\documentclass[AMA,STIX1COL]{WileyNJD-v2}
\usepackage{subfigure}

\articletype{Article Type}%

\begin{document}

\title{Data Poisoning Attacks on Neighborhood-based Recommender Systems}

\author[1,2]{Liang Chen*}

\author[1,2]{Yangjun Xu}

\author[1,2]{Fenfang Xie}

\author[3]{Min Huang}

\author[1,2]{Zibin Zheng}


\address[1]{\orgdiv{The school of data and computer science}, \orgname{Sun Yat-Sen University}, \orgaddress{\state{Guangzhou}, \country{China}}}

\address[2]{\orgdiv{National Engineering Research Center of Digital Life}, \orgname{Sun Yat-sen University}, \orgaddress{\state{Guangzhou}, \country{China}}}

\address[3]{\orgdiv{College of software}, \orgname{South China University of Technology}, \orgaddress{\state{Guangzhou}, \country{China}}}

\corres{*Liang Chen, the school of data and computer science, Sun Yat-Sen University, Guangzhou, China. \email{chenliang6@mail.sysu.edu.cn}}


\abstract[Abstract]{Nowadays, collaborative filtering recommender systems have been widely deployed in many commercial companies to make profit. Neighbourhood-based collaborative filtering is common and effective. To date, despite its effectiveness, there has been little effort to explore their robustness and the impact of data poisoning attacks on their performance. Can the neighbourhood-based recommender systems be easily fooled? To this end, we shed light on the robustness of neighbourhood-based recommender systems and propose a novel data poisoning attack framework encoding the purpose of attack and constraint against them. We firstly illustrate how to calculate the optimal data poisoning attack, namely UNAttack. 
We inject a few well-designed fake users into the recommender systems such that target items will be recommended to as many normal users as possible. Extensive experiments are conducted on three real-world datasets to validate the effectiveness and the transferability of our proposed method. Besides, some interesting phenomenons can be found. For example, 1) neighbourhood-based recommender systems with Euclidean Distance-based similarity have strong robustness. 2) the fake users can be transferred to attack the state-of-the-art collaborative filtering recommender systems such as Neural Collaborative Filtering and Bayesian Personalized Ranking Matrix Factorization.}


\maketitle


\section{Introduction}

With the booming of information, it is a big challenge for users to find valuable items that satisfy their preference. To address this challenge, most of the existing recommender systems studies \cite{covington2016deep,xu2018improving,xie2019generative} utilize users' historical information, social network information and items' content to model users' preference on items. Among these studies, one of the most widely used technique in  recommender systems is collaborative filtering (CF)\cite{chen2019matching}, which focuses on the users' historical information. Depending on the methods utilized to learn the correlation between users' historical information, CF recommender systems can be divided into four categories: matrix-factorization-based methods \cite{Xie2018Factorization}, graph-based methods \cite{xie2018weighted}, association-rule-based methods \cite{davidson2010youtube,mobasher2000automatic} and neighbourhood-based methods \cite{sarwar2001item}. Numerous commercial companies (e.g., Netflix\footnote{https://www.netflix.com}, YouTube\footnote{https://www.youtube.com}, eBay\footnote{https://www.ebay.com} and Taobao\footnote{https://www.taobao.com}) have already applied CF recommender systems to their products, such as web and app, to alleviate the information overload, improve users' experience, and  bring them tremendous economic benefits.

Recommender systems have the advantages of matching users personal interest, however, the overall recommendation result is less robust. This problem is a non-ignorable downside. Several previous studies \cite{o2004collaborative,mobasher2007toward} have already pointed out that CF methods are vulnerable to data poisoning attacks inducing the recommender system to a security risk. The collaborative recommendation system will personalize the recommendation to the user based on the historical data of similar users, when it works normally. However, this may not be the case. For example, an attacker with bad intentions (or a profit motive) quietly injects malicious data with elaborate construction into recommender systems so that he can control the recommendation result as he desires and destory the personal recommendation list.

Early data poisoning studies \cite{lam2004shilling,mobasher2007toward} have applied handcraft rules to generate fake users, which usually achieve suboptimal attack performance even if in the case of possessing the access to input data and knowing the recommendation algorithm. 
For example, random attack\cite{mobasher2007toward} randomly chooses filler-items for each fake user and assigns ratings to the filler-items from the normal distribution of all the rating data. Average attack\cite{mobasher2007toward} selects filler-items just the same as the random attack. The difference is that the rating score assigned to the filler-item is based on the normal distribution of the rating data of the filler-item.
These methods can't test the robust of recommender systems completely. In recent years, several studies \cite{li2016data,yang2017fake,fang2018poisoning} about optimal data poisoning attack against a certain type of recommender systems have been proposed. 
However, how to design an optimal data poisoning attack based on neighborhood-based recommender systems remains a challenging problem.

Despite some advanced recommender systems have been proposed, neighborhood-based collaborative filtering remains one of the most common and effective recommender systems\cite{bell2007improved,ning2015comprehensive,aggarwal2016neighborhood,jannach2017recurrent} and can be deployed by businesses companies, e.g., Amazons \cite{linden2003amazon}. In addition, the recently proposed research \cite{dacrema2019we} points out that neighborhood-based collaborative filtering outperforms than some advanced deep the collaborative filtering models, (e.g., NCF \cite{he2017neural}). 
Thus, proposing an optimal data poisoning attack to test its robustness is an urgent and important issue. With this motivation, we propose an optimal data poisoning attack based on neighborhood-based recommender systems, namely UNAttack. Neighborhood-based recommender systems are mainly based on the similarity between users or items. The user's preference on an item is learned through the user's or the item's $K$ nearest neighbours. In our method, the characteristics of the users' $K$ nearest neighbours will be utilized to promote target items (push attack \cite{mobasher2007toward}). For example, in recommender systems, target items will be recommended to as many normal users as possible after we inject a few fake users into the recommender systems. In data poisoning attack, the items which fake users rate is named as \textbf{filler-items}. Due to the limited resources and to keep our attack hard to be detected, some constraints are defined as follows: 1) in each attack, attackers can only inject $j$ fake users and each fake user can only rate $z$ filler-items at most, 2) the rating scores user give to the items must be integer. We regard data poisoning attack as an optimization problem with the aforementioned constraint in order to get the filler-items and the rating scores of each fake user.  To demonstrate its effectiveness, extensive experiments are conducted on three real-world datasets. The experimental results show that our UNAttack performs far better than conventional attack methods and other state-of-the-art attack methods. In addition, we observe that the fake users we generate can attack the state-of-the-art CF recommender systems, such as NCF\cite{rendle2009bpr} and BPRMF\cite{rendle2009bpr}. It means that UNAttack is still effective even if the attackers do not know the details of the recommended systems, namely the black-box attack \cite{kurakin2016adversarial}.

In summary, the main contributions of this paper are as follows
\begin{itemize}
	\item  We propose a general and mathematical framework for optimal data poisoning attack against CF recommender systems. 
	\item  We encode data poisoning attack against neighborhood-based methods into our framework as an optimization problem. Then, we present the solution to this problem so as to generate more effective fake users.
	\item  To demonstrate how our method works, we conduct a series of experiments on three public datasets. Both quantitative and qualitative analysis justify that our attack methods can achieve good attack performance, not only in neighborhood-based recommender systems, but also in other collaborative filtering algorithms such as BPRMF and NCF.
\end{itemize}

\section{Related work}

The existing research about data poisoning attack has attracted the attention of many people. \cite{alfeld2016data} propose a data poisoning attack against autoregressive models using the optimal methods. \cite{zhang2019data} introduce data poisoning strategies to test knowledge graph embedding robustness. \cite{ma2019data} design  data poisoning attack algorithms targeting objective and output. Existing research about data poisoning attack (also called as shilling attacks\cite{lam2004shilling}) against the recommender system studied from more than 10 years ago. Data poisoning attacks aim to make target items be recommended to as more users as possible. Specifically, when performing data poisoning attacks, attackers firstly registers a few fake accounts in the service. Then they control each fake account to assign well-esigned rating scores to a carefully chosen subset of items. 
\cite{lam2004shilling} proposed two kinds of data poisoning attack methods: random attack and average attack. Both these attack methods randomly chose filler-items for each fake user.
In \cite{mobasher2007toward}, the authors proposed the bandwagon attack, which associated the filler-items with popularity. Different from the random attack, bandwagon attack selected a part of the popular items together with some randomly selected items as filler-items.

In recent years, a few data poisoning attack methods, which uses the optimization technique to get the filler-items and the assigned ratings of fake users, have been proposed. A study \cite{li2016data} proposed a data poisoning attack against matrix-factorization-based recommender systems making the root-mean-squared-error become larger than its original value. Fake co-visitation\cite{yang2017fake}is a injection attack against association-rule-based recommender systems by constructing attack as a constrained linear optimization problem. \cite{fang2018poisoning} proposed an optimized data poisoning attack for graph-based recommender systems. 

Besides, the paper \cite{xing2013take} proposed a new type of attack method named profile pollution attack against recommender system and other web services, in which attackers aimed to pollute users' profile, such as browsing and clicking session, via cross-site request forgery \cite{zeller2008cross}. However, they only can perform the attack on a small scale. 
In recently, leakage of private information has attracted more and more people’s attention\cite{Yi2015Security} and especially the edge computing and fog computing are used, which can collect more sensitive information than the remote cloud.
Privacy attack mainly contains the item inference attacks and attribute inference attacks.
 The work \cite{calandrino2011you} proposed privacy attacks to infer the items that a target user has rated via utilizing the publicly available reviews of users. The authors have already proved that these methods can be performed on several popular webs such as Amazon\footnote{https://www.amazon.com/}, LibraryThing\footnote{https://www.librarything.com/} and Lastfm\footnote{https://www.last.fm/zh/}. Attribute inference attacks \cite{gong2016you} are the technique aiming to acquire the users' privacy attributes with the users' interaction information. Generally, users' privacy attributes (e.g., gender, political view, interests, and location) will be reflected by the users' interaction information (e.g., buy, rate and click items). After collecting a few users' interaction information and their privacy attributes, the attacker trained a classifier model with interaction information as input and privacy attributes as the prediction. This classifier model will be applied to predict the attributes of users who are unwilling to make their privacy attributes public.
These methods have been demonstrated to be feasible by several studies\cite{gong2016you,gong2018attribute,gong2014joint}.

\section{Method}
We first introduce the formulation of neighborhood-based recommender systems. Then, we define data poisoning attacks as an optimization problem and approximate this optimization problem. Lastly, the solution to this problem will be introduced to generate fake users.

\begin{figure*}[t]
	\centering
	\begin{minipage}{3in}
		\includegraphics[width=3.2in,height=2.8in]{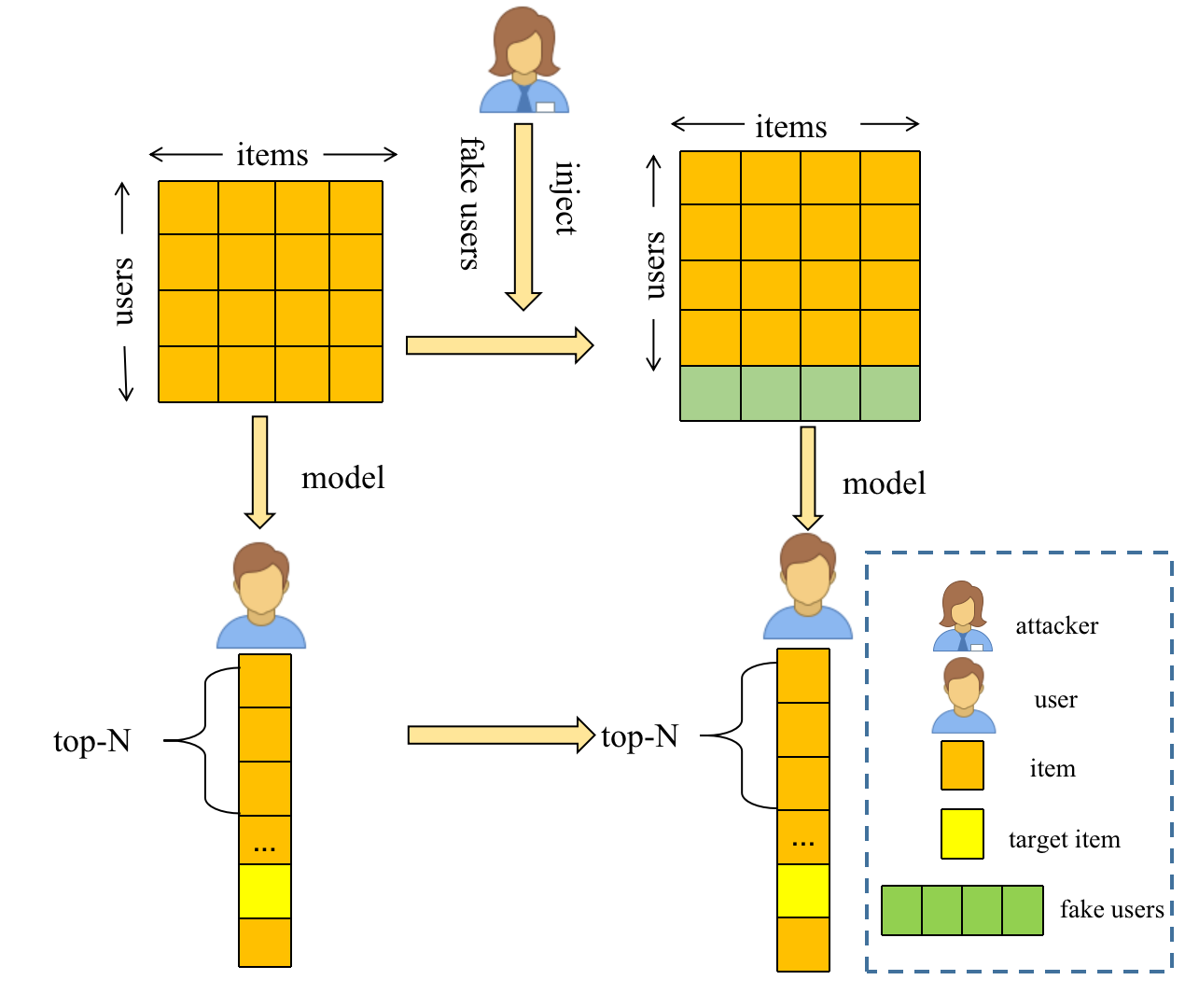}%
	\end{minipage}
	\caption{intuitive sketch to explain the process of data poisoning.} \label{fig0}
\end{figure*}

\subsection{Neighborhood-based recommender systems}
Neighborhood-based recommender systems are the earliest CF recommender systems method including user-based CF and item-based CF. Both of them have been widely applied in various web services. In the user-based CF, when the model recommends items to the user, it selects the top-K nearest-neighbors of the user via calculating the similarity between the user and others, and then predicts the preference of the user on items based on the preference of top-K nearest-neighbours. For the item-based CF, the model considers the similarity based on the items. In this paper, we mainly focus on using our methods against user-based CF by injecting fake users into the recommender systems.
We assume there is a set of $m$ users, $U = \{u_{1},u_{2},...,u_{m} \}$, a set of $n$ items, $I =  \{i_{1},i_{1},.....,i_{n} \}$ and a sparse matrix $R_{m \times n}$. $X_{u}$ represents the n-dimensional item rating vector for user $u$. Each element $X_{ui}$ in the vector $X_{u}$ denotes the explicit preference of user $u$ on item $i$. The predicted preference of user $u$ on item $i$ based on user-based CF can be calculated as follows:
\begin{equation}
p_{ui} = \sum _{v\in S(u,K)\cap U_{i}^{+})} s_{uv}X_{vi}
\end{equation}
where $S(u,K)$ is the set of  top-K nearest neighbours of user $u$, $U_{i}^{+}$ denotes a set of users who rate item $i$. $s_{uv} $ is the similarity between user $u$ and $v$.

\subsection{Problem Definition}
According to the intent, data poisoning attack can be divided into two categories: target attacks and non-target attacks \cite{mobasher2007toward}. In non-target attacks, the intent could be to deteriorate the recommendation quality of all users. \cite{li2016data} proposed a data poisoning attack against matrix-factorization-based recommender systems making the root-mean-squared-error become larger than its original value. Our work will focus on the target attack. We consider the case where an attacker generates a few of realistic-looking fake users with the desires of promoting a group of items from himself company into target users' top lists or removing a group of items from competitor company from target users' top lists. Figure \ref{fig0} illustrates how data poisoning attacks work against the recommender systems. 
Thus, the performance of our data poisoning attack method can be measured via $HitRatio$, and the $HitRatio$ of item $i$ is defined as follows:
\begin{equation}
\begin{split}
H_{ui}=\left\{
\begin{aligned}
1 , && i \ \epsilon\  L_{u} \\
0 ,&& otherwise
\end{aligned}
\right.\\
\\
HitRatio_{i} = \frac{\sum_{u\in  U_{i}^{-} } H_{ui}}{\left | U_{i}^{-} \right | }\\
\end{split}	
\end{equation}	

Let $L_{u}$ denote the set of top-N items list that user-based CF recommend to user $u$. $U_{i}^{-}$ denotes a set of users who do not rate item $i$, namely target users, and $| U_{i}^{-} |$ is the number of target users.

Considering the attacker's intent in the paper is to promote target items into as many target users' top-N lists as possible. Demoting target items out of target users' top-N lists is the special case of promoting\cite{mobasher2007toward}. For concreteness, attackers can promote other items so as to push target items out of target users' top-N lists.
Therefore, the key challenge for us is how to generate an n-dimensional item rating vector for each fake user who can maximize the $HitRatio$ of the target item. In reality, attackers often suffer from some constraints in poisoning the data such as the cost of generating fake users and detection avoidance. As such, the number of fake users, the number of filler-items of each fake user and the range of rating scores will be restricted. Let $X_{f}$ be the n-dimensional item rating vector for fake user $f$, in which $X_{fi}$ is the score that fake user $f$ gives to the item $i$. Under the consideration above, we describe data poisoning attack task as the following optimization problem:

\begin{equation}
\begin{split}
Maximize(HitRatio)\\
s.t. \left | X_{f} \right | ^{'} \leq z, \\
X_{fi} \in \left \{0, 1, ..., r_{max} \right \},\\
\forall f \in \left \{ f_{1},f_{2},...,f_{j} \right \}\\
\end{split}	
\end{equation}
where $ \{ f_{1},f_{2},...,f_{j} \} $ denotes the set of $j$ fake users, $|X_{f}|^{'}$ denotes the number of filler-items for each fake user ($|X_{f}|^{'} \leq  z$). The rating scores of items are in the interval $\{0, 1, ..., r_{max} \}$. It is worth noting that this framework in Eq.3 is suitable for target data poisoning attack against any recommender systems. In this paper,  we focus on encoding the neighborhood-based recommendation into this framework.

\subsection{Problem Approximation}
It is clearly intractable to find the optimal solution to the optimized problem in Eq.3 due to the following two reasons. Firstly, the item rating vector $X_{f} (f \in \{ f_{1},f_{2},...,f_{j} \})$ of the fake user has the implicit and complicated dependency with $HitRatio$. Secondly, the rating score $X_{fi}$ is a discrete value, which means that it can not be optimized via gradient-based methods such as gradient descent.

To overcome the difficulties mentioned above, several approximation techniques have been proposed. Firstly, we will optimize the fake users one by one based on the new data (including the normal users and the fake users generated before). Secondly, we borrow the strategy from the ranking problem to construct pairwise loss function, in which after training item with a higher value means the user is more likely to buy. Through this way, we obtain the filler-items of the fake user by selecting top-$z$ items with the highest scores after training. Thirdly, the filler-items will be assigned integer rating scores to mimic normal user behaviours.

If attackers want to maximize the $HitRatio$, he should make the target item $t$ appear in the top-N recommended list $L_u$ as more as possible for each user $u \in U_{t}^{-}$. To achieve this purpose, the objective function must be converted into the loss function that is convenient to be optimized. For user $u$, if $t$ appears in the recommended list $L_{u}$, then users' rating scores to $t$ must be higher than item $i$ (${i \in L_{u}}$), which can be expressed as $ p_{ut} > p_{ui}$. However, according to Eq.1, we can know that the user-based CF evaluates the user's preference through his top-K nearest neighbors' preference. Therefore, if the attacker wants to impose influence on $p_{ut}$ and $p_{ui}$, he must make the fake user $f$ be in the top-K nearest neighbours of user $u$, which can be expressed as $s_{uf} > s_{uv}$. 

More specifically, to approximate the objective function $HitRatio$, the loss function  must satisfy a condition when it becomes smaller, $s_{uf}$ and $p_{ut}$ become higher than $s_{uv}$ and $p_{ui}$, respectively. We borrow the idea from BPR pairwise loss and formulate the user $u$'s loss function as follows:
\begin{equation}
\begin{split}
loss_{1} &= \sum_{v\in S(u,K)}\sigma (s_{uv} - s_{uf})\\
loss_{2} &= \sum_{i\in L_{u}}\sigma (p_{ui} - p_{ut})\\
loss_{u} &=  (1- \lambda)loss_{1} + \lambda loss_{2}\\
\end{split}
\end{equation}
where $ \sigma (x) = \frac{1}{1 + e^{-x}}$ is the sigmoid function. $\lambda \in \left [ 0,1 \right ]$ is the trade-off parameters. $\lambda$ controls the relative importance of $ loss_{1}$  and $loss_{2} $. By minimizing $loss_{u}$, we can promote the fake user into the top-K nearest neighbours $S(u,K)$ and promote the target item into the user's top-N items $L_{u}$ list.

For all the normal users in $U_{t}^{-}$, the loss function is the sum of their single loss function:
\begin{equation}
loss  = \sum_{u\in U_{t}^{-}}loss_{u}
\end{equation}

With the loss function, we can consider the optimization problem as follows:
\begin{equation}
\begin{split}
&Minimize(F(X_{f})=loss)\\
&s.t. \left | X_{f} \right | ^{'} \leq z,\\
&X_{fi} \in \left \{0, 1, ..., r_{max} \right \}\\
\end{split}
\end{equation}

\subsection{Fake Users Generation}

We elaborate the following specific steps on how to solve the optimized problem in Eq.6 and generate fake users.\\
\setlength{\parindent}{0em}
\textbf{1) choosing the optimal filler-items for fake users.} In this step,  the stochastic gradient descent will be chosen to solve the problem in Eq.6 due to its ease in deriving the update strategy. Discrete value is hard to be optimized by gradient descent, so $X_{fi}$ is relaxed as a 
continuous value in training phrase. In this section,we choose the Cosine Similarity to calculate the similarity between users as the illustrative example. In iteration $t$,  $X_f(t)$ will be updated as follows:
\begin{equation}
X_{f}^{(t)} = Project(X_{f}^{(t-1)}-\eta \frac{\partial F(X_{f})}{\partial X_{f}} )\\
\end{equation}
where $Project(x)$ is the project function that cuts each $X_{fi}$ into the range  $[0,1,..r_{max}]$. The gradient of $F(X_{f})$ (with respect to  $X_{f}$) are as follows:
\begin{equation}
\begin{split}
\frac{\partial F(X_{f})}{\partial X_{f}}  =  \sum_{u\in U_{t}^{-}}(1-\lambda) \frac{\partial loss_{1}}{\partial X_{f}} + \lambda \frac{\partial loss_{2}}{\partial X_{f}}
\end{split}
\end{equation}
\begin{equation}
\begin{split}
\frac{\partial (loss_{1})}{\partial X_{f}}
& = \sum_{v\in S(u,k) }\frac{\partial \sigma (Q)}{\partial Q}(\frac{\partial s_{uv}}{\partial X_{f}} - \frac{\partial s_{uf}}{\partial X_{f}})
\end{split}
\end{equation}
\begin{equation}
\begin{split}
\frac{\partial (loss_{2})}{\partial X_{f}}
& = \sum_{i\in L_{u} } \sum_{v\in W} \frac{\partial \sigma (P)}{\partial P}(\frac{\partial s_{uv}X_{vi}}{\partial X_{f}} - \frac{\partial s_{uf}X_{ft}}{\partial X_{f}})
\end{split}
\end{equation}\\
where $Q = s_{uv} - s_{uf} $, $P = p_{ui} - p_{ut}$ and $ W =  (S(u,k)\cap U_{i}^{+})$

\setlength{\parindent}{2em}
Before calculating the $\frac{\partial F(X_{f})}{\partial X_{f}}$, we must solve the $\frac{\partial s_{uf}}{\partial X_{f}}$. $s_{uf}$ denotes the similarity between user $u$ and $f$. In Eq.9 and 10, the Cosine Similarity will be used and the gradient  $\frac{\partial s_{uf}}{\partial X_{f}}$ can be computed as follows:

\begin{equation}
\begin{split}
\frac{\partial s_{uf}}{\partial X_{f}} =\frac{X_{u}}{\left \| X_{u} \right \| \left \| X_{f} \right \|} - \frac{{}X_{u}X_{f}}{\left \| X_u \right \|\left \| X_{f} \right \|}\frac{X_{f}}{\left \| X_{f} \right \|^{2}}\\
\end{split}
\end{equation}\\
\setlength{\parindent}{0em}
\textbf{2) assigning integer rating scores to filler-items.} After finishing the training of $X_{f}$, we use the following several strategies to convert the continuous value into integer rating score: first of all, let $X_{ft} = r_{max}$. In other words, we give the target items the maximum ratings. Secondly, inspired by the ranking problem, all items will be ranked according to $X_{fi}$, and top-$z$ items with the highest values will be chosen as the filler-items. Because we consider that after training the items with higher value means the fake user is more likely to buy them. Finally, the rating score assigned to each filler-item is drawn from a normal distribution of the normal users' rating data of this item. After that, we can keep our fake user mimic the normal user's behaviours as well as avoid potential detection. Algorithm 1 shows the detailed procedure of our solution. The time cost of fake users generation is  $j \times z$  , where $j$ is the number of fake users and $z$ is the number of filler items.

\begin{algorithm}[!htbp]
	\caption{UNAttack}
	\label{alg:algorithm}
	\textbf{Input}: Matrix $R_{m\times n}$\\
	\textbf{Parameter}: $\lambda, K, N, z, j$\\
	\textbf{Output}: $j$ fake users 
	\begin{algorithmic}[1] 
		\For {each fake user f}
		\State Solve the problem in Eq.6 with current rating matrix $R$ to get $X_{f}$
		\State  Let $X_{ft} = r_{max}$
		\State  Select $z$ items with highest value in $X_{fi}$ as filler items.
		\State  For each filler-items j, $X_{fj} \sim \mathcal{N}(\mu _{j},\sigma _{j}^{2})$
		\State  $R_{m\times n}= R_{m\times n} \cup X_{f}$
		\EndFor
	\end{algorithmic}
\end{algorithm}

\section{Experiments}
In this section, we empirically evaluate the performance of our proposed method, UNAttack, with all the baselines on three real-world datasets. Then, we explore the impact of hyper-parameters on the UNAttack. Finally, the performance of transferability on the UNAttack will be introduced.
\subsection{Experiment Settings}

\begin{table}[!htbp]
	\caption{Statistics of the datasets.}
	\setlength{\tabcolsep}{1.8mm}
	\centering
	\begin{tabular}{@{}c|ccccc@{}}
		\toprule
		Datasets  & \#users & \#items & \#ratings & sparsity & average \\ \midrule
		FilmTurst & 1,508    & 2,071    & 35,497     & 98.86\%  & 23                    \\
		Movielens & 943     & 1,682    & 100,000    & 95.72\%  & 106                   \\
		Amazons   & 2,378    & 2,525    & 23,613     & 99.60\%  & 10                    \\ \bottomrule
	\end{tabular}
	\label{tab1}
\end{table}

\textbf{Dataset description.} We conduct extensive experiments on three real-world datasets: FilmTrust \cite{guo2013novel}, Movielens\footnote{https://grouplens.org/datasets/movielens/}, and Amazons Video\footnote{http://jmcauley.ucsd.edu/data/amazon/}. Table \ref{tab1} summarizes their detailed statistics. For each dataset, 80\% of interactions information between users and items will be split into the training set and the remaining 20\% as the test set. Then, we randomly select aside 10\% of the training data as validation set for tuning hyper-parameters of the recommendation model. The word \textbf{average} in the Table \ref{tab1} means the average rating number of each users. The definition of sparsity in Table \ref{tab1} is as follows:
\begin{equation}
Sparsity = 1 - \frac{\#ratings}{\#users \times \#items}
\end{equation}

\setlength{\parindent}{0em}
\textbf{Compared methods.} We compared our method with several data poisoning attack methods using the $HitRatio@N$ ($HR@N$).

\begin{itemize}
	\item \textbf{None.} This represents the situation that recommendation is normal and do not suffer any attack.
	\item \textbf{Random attack \cite{lam2004shilling}.} This method randomly chooses filler-items for each fake user, and assigns ratings to the filler-items from the normal distribution of all the rating data.
	\item \textbf{Average attack \cite{lam2004shilling}.} It selects filler-items just the same as the random attack. However, the rating score assigned to the filler-item is based on the normal distribution of the rating data of the filler-item.
	\item \textbf{Bandwagon attack \cite{mobasher2007toward}.} Bandwagon attack associates the filler-items with popularity. In our experiments, for each fake user, we select $z \times 20\%$ items which get high average scores and randomly choose $z \times 80\%$ as filler-items, and assign rating to the filler-items from the normal distribution of the whole rating data.
	\item \textbf{Co-visitation attack \cite{yang2017fake}.} Fake Co-visitation injection attack is designed for the association-rule-based recommender systems. Our experiment only considers the attack with injecting fake users. Therefore, we use the method in Fake Co-visitation injection attack to choose the filler-items for fake users. Moreover, if the item $i$ is frequently rated with the target item $t$ at the same time, it has a high probability to be selected as filler-items. The way we assign the rating score for filler-items is the same as the average attack.
\end{itemize}

\setlength{\parindent}{0em}
\textbf{Parameter Setting.} The details of our parameter are as follows.  The best $\lambda$ values will be used for each dataset. $z$  will be set to the average rating number of normal users. Without specification, in FilmTrust, $\lambda = 0.6$, $K = 30$, $N =20$, $z = 23$. In Movielens, $\lambda$ = 0.5, $K = 30$, $N =20$, $z = 106$. In Amazons, $\lambda = 0.3$, $K = 30$, $N =20$, $z = 10$. Besides, the number of fake users (attack size) is set from $0.5\%$ to $2\%$ of the number of normal users. By default, we assume that the similarity between users is calculated by Cosine Similarity. To demonstrate the performance of our method, a part of items are first selected randomly as random  items. Secondly, we consider the items whose rating data is less than 5 and which are almost not recommended to normal users at all (None performance less than 0.001) as cold-start target items. Moreover,  the items with high possibility are recommended to normal users before the attack will be regarded (None performance higher than 0.1) as warm-start items. We select 10 target items for each type of target items (eg. random items, cold-start items and warm-start items) to calculate the average $HR@N$.  All of the methods run on an Intel Core i7 with 2.2 GHz , 2080 Ti GPU,128GB RAM, 64 bit system.

\begin{table*}[]
	\setlength{\tabcolsep}{2.5mm}
	\centering
	\caption{Overall performance between our method and other data poisoning attack methods.}
	
	\begin{tabular}{|l|l||c|c|c|c|c|c|c|c|c|}
		\hline
		\multirow{2}{*}{Dataset}   & \multirow{2}{*}{Method} & \multicolumn{3}{c|}{Random target items}            & \multicolumn{3}{c|}{Cold-start target items}        & \multicolumn{3}{c|}{Warm-start target items}         \\ \cline{3-11} 
		&                         & 0.50\%          & 1\%             & 2\%             & 0.50\%          & 1\%             & 2\%               & 0.50\%          & 1\%             & 2\%             \\ \hline  \hline
		\multirow{6}{*}{FilmTrust} & None                    & 0.0058          & 0.0058          & 0.0058          & 0.0001          & 0.0001          & 0.0001          & 0.3997          & 0.3997          & 0.3997          \\ \cline{2-11} 
		& Random                  & 0.0075          & 0.0087          & 0.0107          & 0.0010          & 0.0013          & 0.0042          & 0.4016          & 0.4033          & 0.4084          \\ \cline{2-11} 
		& Average                 & 0.0075          & 0.0085          & 0.0106          & 0.0006          & 0.0015          & 0.0045          & 0.4017          & 0.4051          & 0.411           \\ \cline{2-11} 
		& Popular                 & 0.0069          & 0.0082          & 0.0101          & 0.0009          & 0.0017          & 0.0035          & 0.4036          & 0.4051          & 0.4084          \\ \cline{2-11} 
		& Co-visitation             & 0.0073          & 0.0102          & 0.0314          & 0.0001          & 0.0022          & 0.0207          & 0.4125          & 0.4182          & 0.4246          \\ \cline{2-11} 
		& UNAttack                & \textbf{0.0578} & \textbf{0.0881} & \textbf{0.1274} & \textbf{0.0533} & \textbf{0.0827} & \textbf{0.1142} & \textbf{0.4566} & \textbf{0.4804} & \textbf{0.4984} \\ \hline
		\hline
		\multirow{6}{*}{Movielens} & None                    & 0.0057          & 0.0057          & 0.0057          & 0.0000          & 0.0000          & 0.0000          & 0.2472          & 0.2472  
		& 0.2472          \\
		\cline{2-11} 
		& Random                  & 0.0057          & 0.0057          & 0.0057          & 0.0000          & 0.0000          & 0.0000          & 0.2472          & 0.2472          & 0.2472          \\ \cline{2-11} 
		& Average                 & 0.0057          & 0.0057          & 0.0057          & 0.0000          & 0.0000          & 0.0000          & 0.2472          & 0.2472          & 0.2472          \\ \cline{2-11} 
		& Popular                 & 0.0057          & 0.0057          & 0.0057          & 0.0000          & 0.0000          & 0.0000          & 0.2472          & 0.2472          & 0.2472          \\ \cline{2-11} 
		& Co-visitation               & 0.0062          & 0.0107          & 0.0469          & 0.0000          & 0.0004          & 0.0263          & 0.2556          & 0.2700          & 0.3103          \\ \cline{2-11} 
		& UNAttack                & \textbf{0.0093} & \textbf{0.0202} & \textbf{0.0911} & \textbf{0.0000} & \textbf{0.0040} & \textbf{0.0675} & \textbf{0.2655} & \textbf{0.2970} & \textbf{0.4142} \\ \hline
		\hline
		\multirow{6}{*}{Aamzons}   & None                    & 0.0036          & 0.0036          & 0.0036          & 0.0004          & 0.0004          & 0.0004          & 0.1897          & 0.1897          & 0.1897          \\ \cline{2-11} 
		& Random                  & 0.0097          & 0.0184          & 0.0429          & 0.0058          & 0.0151          & 0.039           & 0.2055          & 0.2186          & 0.2525          \\ \cline{2-11} 
		& Average                 & 0.0131          & 0.023           & 0.0485          & 0.0085          & 0.0175          & 0.0432          & 0.2052          & 0.2207          & 0.2527          \\ \cline{2-11} 
		& Popular                 & 0.0126          & 0.0213          & 0.0452          & 0.0074          & 0.0161          & 0.0362          & 0.2044          & 0.2167          & 0.2445          \\ \cline{2-11} 
		& Co-visitation             & 0.0199          & 0.0437          & 0.1035          & 0.0154          & 0.0400          & 0.1015          & 0.2254          & 0.3004          & 0.3194          \\ \cline{2-11} 
		& UNAttack                & \textbf{0.0912} & \textbf{0.1304} & \textbf{0.1794} & \textbf{0.0926} & \textbf{0.1209} & \textbf{0.1572} & \textbf{0.2837} & \textbf{0.309}  & \textbf{0.3725} \\ \hline
	\end{tabular}
	\label{tab2}
\end{table*}

\begin{figure*}[t]
	\centering
	\subfigure[]{
		\begin{minipage}{2.2in}
			\centering
			\includegraphics[width=2.2in,height=1.6in]{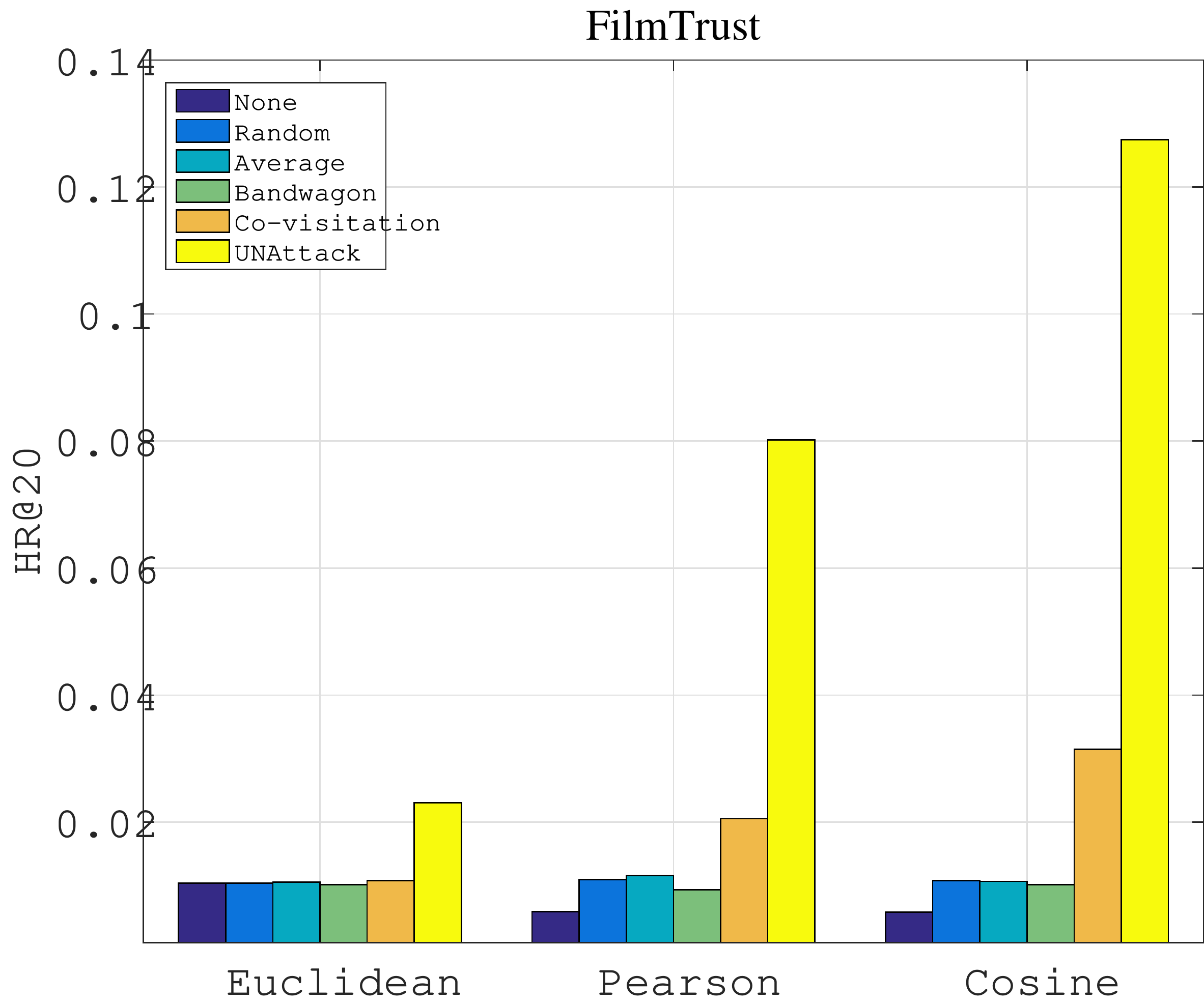}%
		\end{minipage}
	}
	\subfigure[]{
		\begin{minipage}{2.2in}
			\centering
			\includegraphics[width=2.2in,height=1.6in]{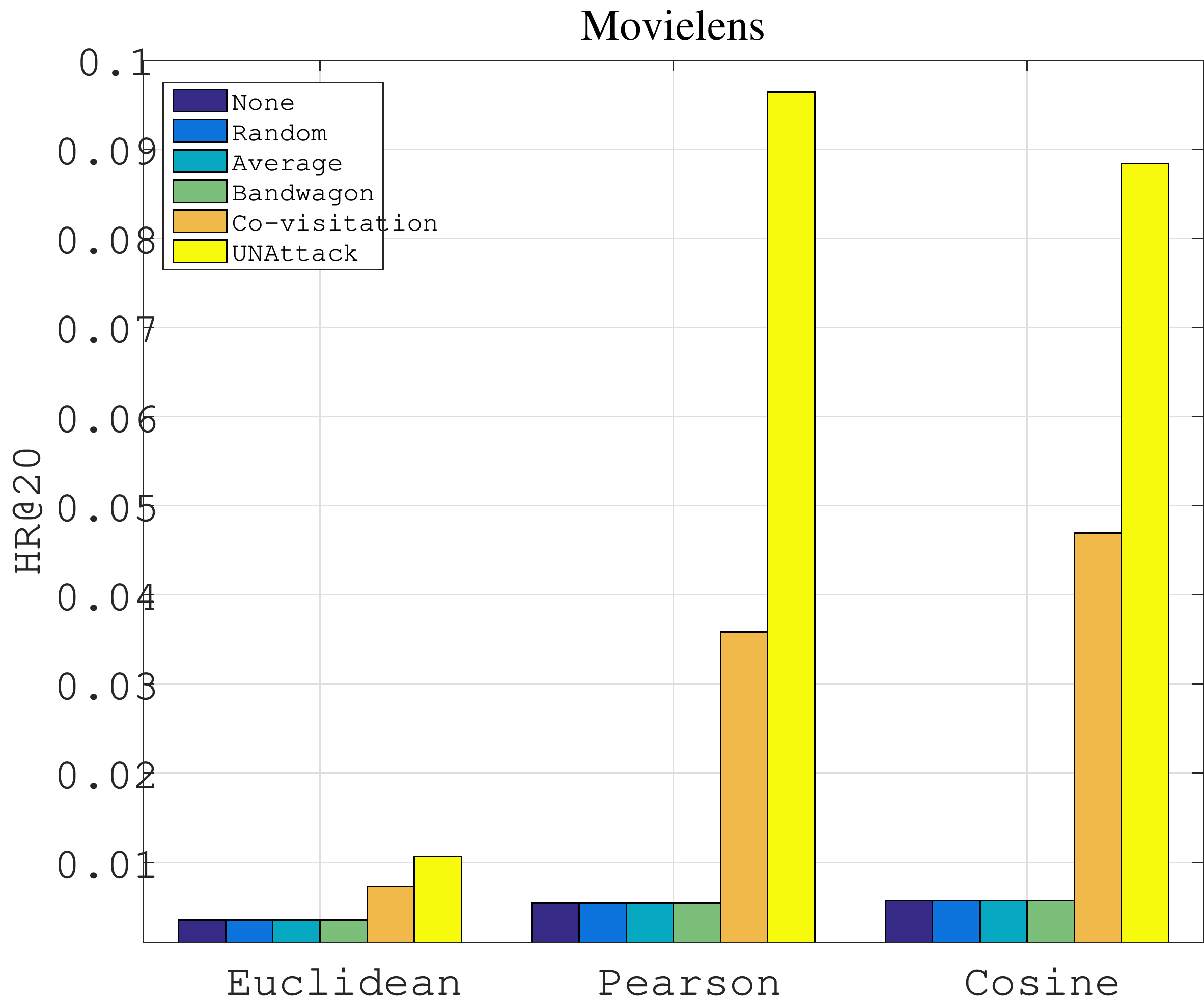}%
		\end{minipage}
	}
	\subfigure[]{
		\begin{minipage}{2.2in}
			\centering
			\includegraphics[width=2.15in,height=1.6in]{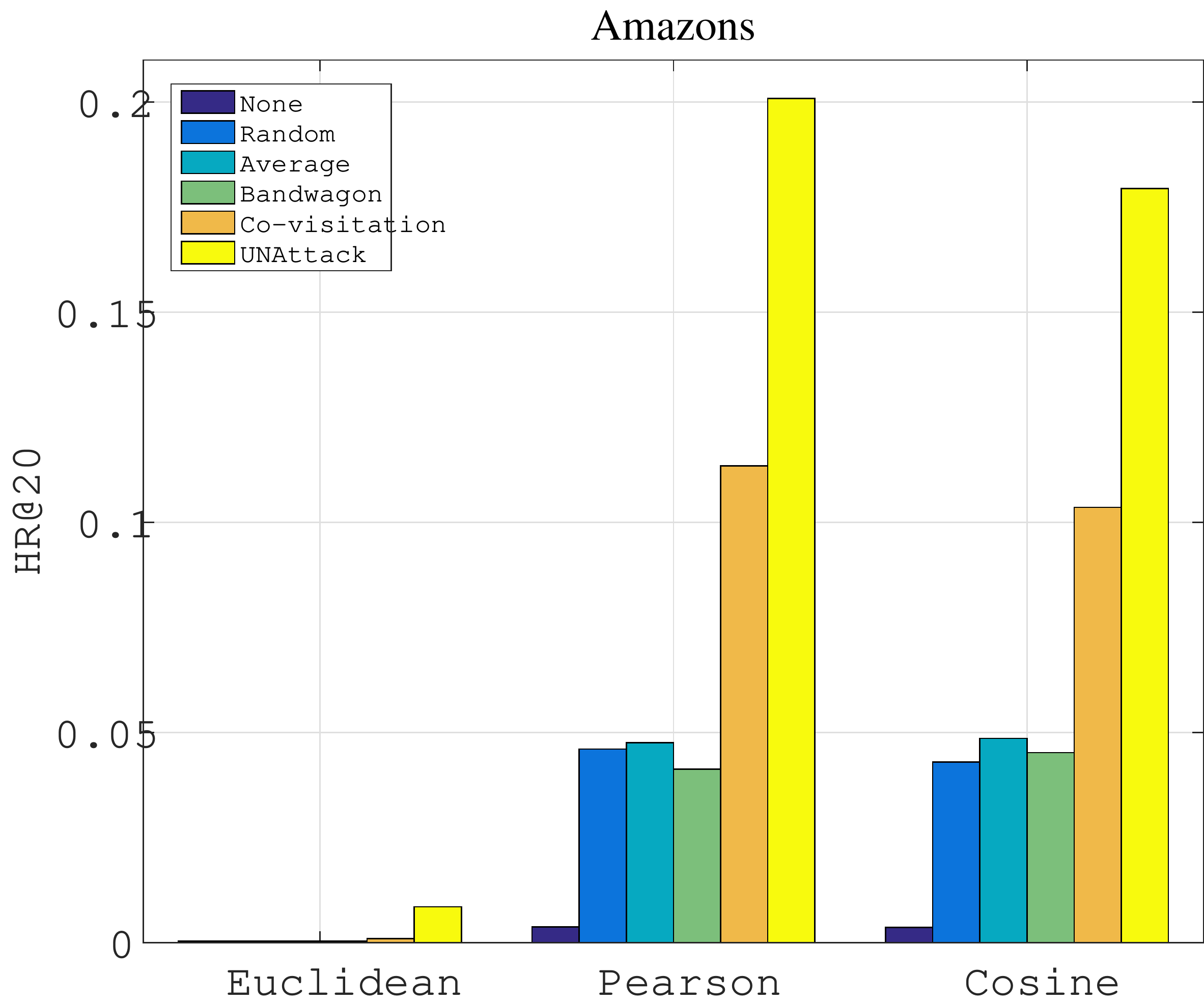}%
		\end{minipage}
	}
	\caption{Impact of the type of similarity measurement method.} \label{fig1}
\end{figure*}

\subsection{Experiment Results}

\textbf{Overall performance.} Table \ref{tab2} summarizes the overall performance of baselines and our attack method. From the table, we can obtain the following observations:
\begin{itemize}
	\item Our attack method can promote the $HitRatio$ in different types of target items effectively by injecting a few fake users. For example, when injecting $1\%$ fake users, in FilmTrust datasets, our attack method increases the $HitRatio$ 15.2 times as much as None in random target items.  In Amazons dataset, the $HitRatio$ of our attack method is 35.9 times higher than None in random target items. For the warm-start items, our method improves $HitRatio$ by $20.1\%$ than None in the Filmtrust dataset.
	\item our UNAttack achieves the highest threat in the same attack size comparing with other methods. For example, in FilmTrust datasets, when injecting $1\%$ fake users, the $HitRatio$ of our attack method attacks random target items is 10 times higher than Random attack, Average attack, and Bandwagon attack, and 8.6 times than Co-visitation attack. If we attack warm-start items, our method can increase the $HitRatio$ by 20.1\%, and the Co-visitation method can only increase the $HitRatio$ by 4.4\%.
	\item The improvement of $HitRatio$ when attacking the cold-start target items is more obvious than random target items. For example, when injecting $1\%$ fake users, in FilmTrust datasets, the $HitRatio$ of our attack method is 15.2 times and 827 times higher than None in random target items and cold-start target items, respectively. In Amazons dataset, our attack method increases the $HitRatio$ 35.9 times and 287 times as compared with None in random target items and cold-start target items, respectively. In addition, the increment of $HitRatio$ in warm-start target items is less effective than in random target items. For instance, the $HitRatio$ of UNAttack is 1.3 and 15.2 times higher than None in warm-start target items and random target items, respectively.
	\item Data poisoning attack can achieve better performance in the sparse dataset than the dense dataset. For example, when injecting $2\%$ fake users and attacking the random target items, in Amazons dataset, the $HitRatio$ of our attack method achieves the highest performance (0.1794), and in the dense dataset, Movielens, the $HitRatio$ is the lowest (0.0911).
	
	\item We can find that the time cost of fake users generation with UNAttack is small. In the Movielens dataset, attackers only spend 8m56s if they generate the number of fake users equivalent to 2\% of the number of normal users. In the Filmtrust and Amazons dataset, they spend 12m8s and 28m27s respectively.
	
\end{itemize}

\textbf{Impact of the type of similarity measurement method.} User-based CF can take different types of similarity measurement methods to calculate the similarity between users. Figure \ref{fig1} shows the result of all attack methods against the User-based CF with different similarity measurements, where the attack size = $2\%$ and we attack random target items. The similarity measurement methods includes Cosine Similarity, Euclidean Distance-based Similarity and Pearson Similarity. We can observe an interesting insight that the performance of all attack methods on Cosine Similarity and Pearson Similarity is better than that on Euclidean Distance-based Similarity. The reason is that Euclidean Distance-based Similarity focuses on the distance of two vectors in space, while Cosine and Pearson Similarity focus on the direction of the two vectors in space. The experimental results demonstrate that fake users are more similar in direction with real users and it is hard to reduce the distance between them. In other words, it is difficult for attack method to promote $HitRatio$ if user-based CF uses the Euclidean Distance-based similarity. Therefore, the user-based CF with Euclidean Distance-based similarity has better robustness. 

\begin{figure*}[!ht]
	\centering
	\subfigure[]{
		\begin{minipage}{2.2in}
			\centering
			\includegraphics[width=2.2in,height=1.6in]{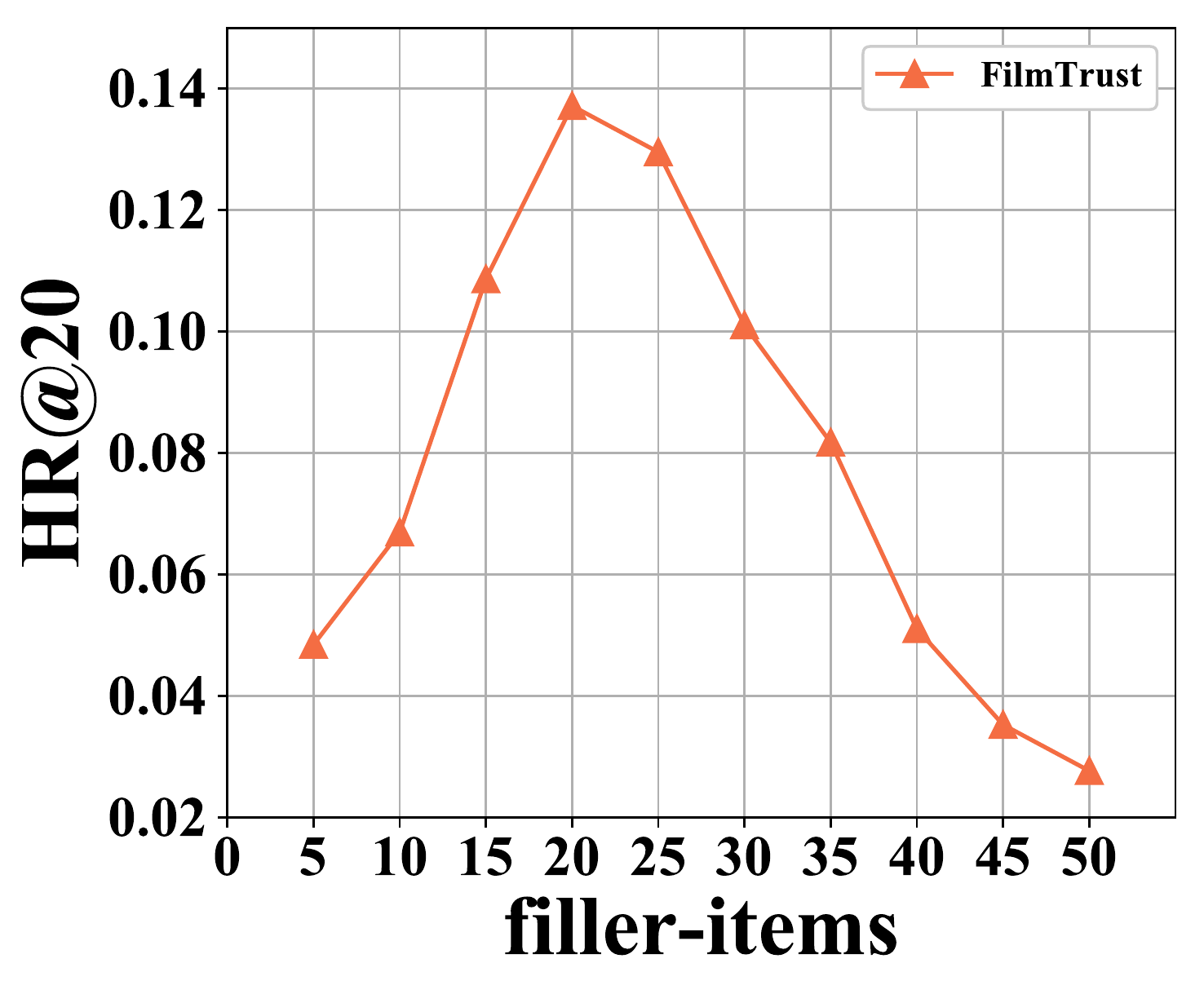}%
		\end{minipage}
	}
	\subfigure[]{
		\begin{minipage}{2.2in}
			\centering
			\includegraphics[width=2.2in,height=1.6in]{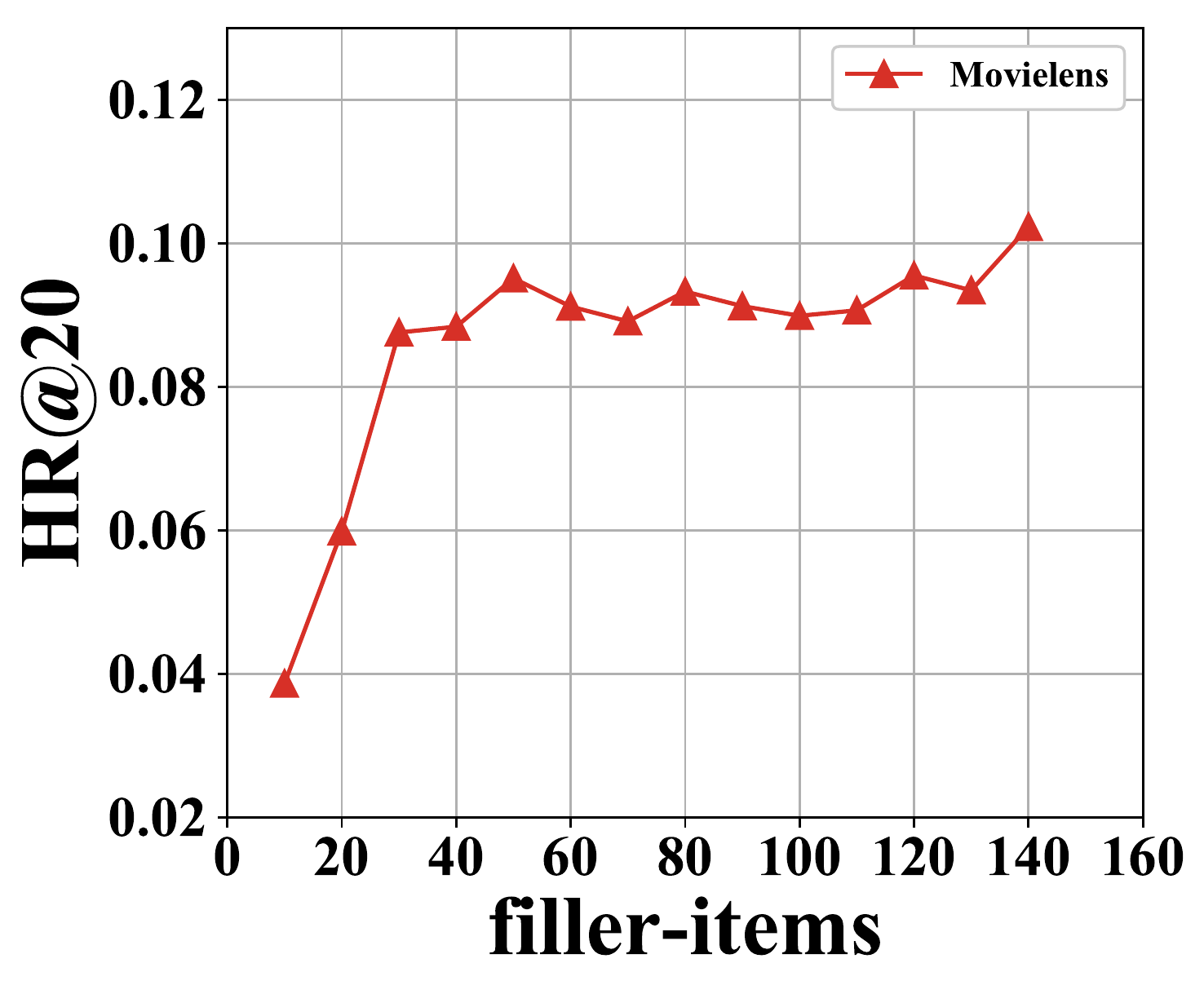}%
		\end{minipage}
	}
	\subfigure[]{
		\begin{minipage}{2.2in}
			\centering
			\includegraphics[width=2.2in,height=1.6in]{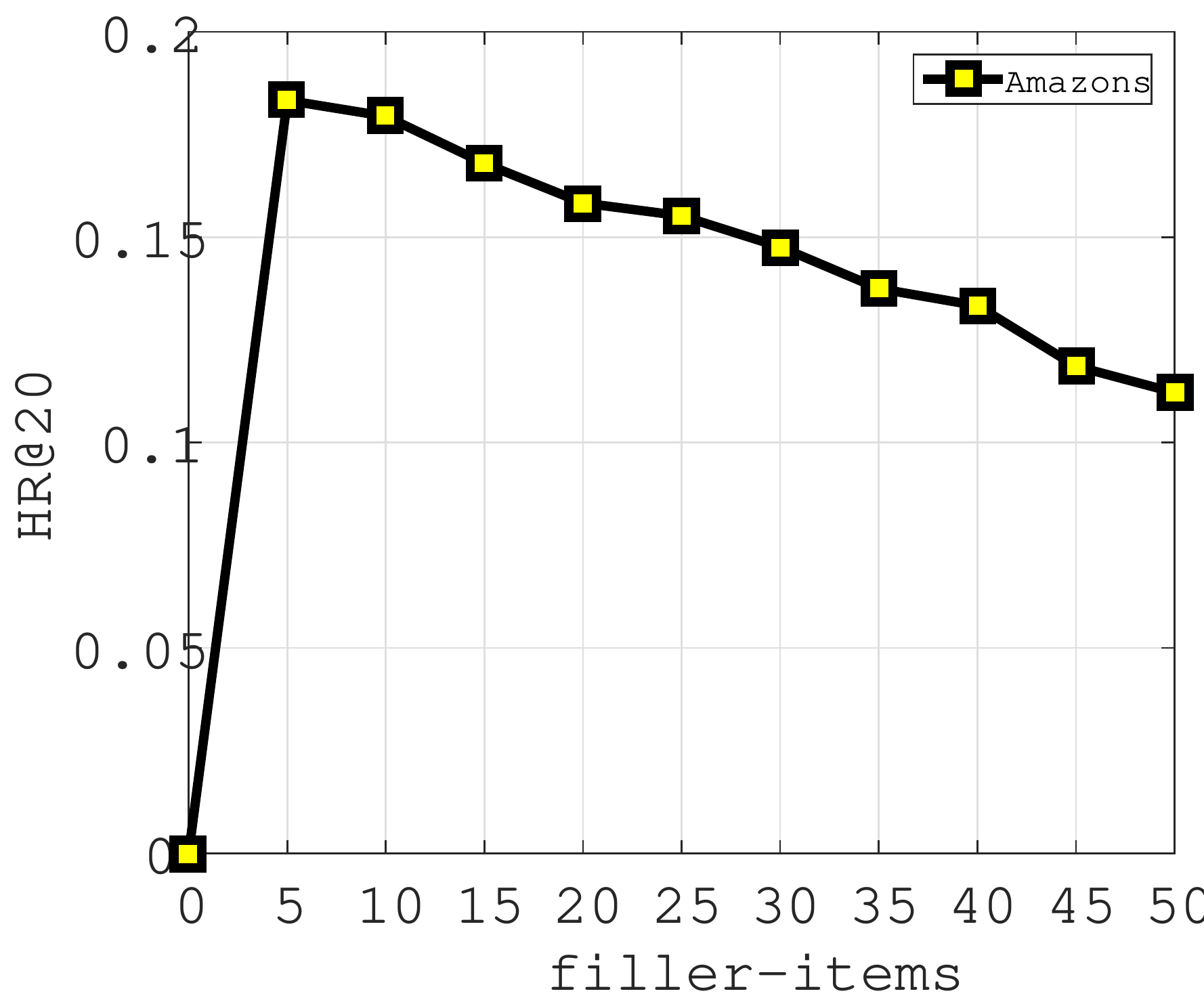}%
		\end{minipage}
	}
	
	\subfigure[]{
		\begin{minipage}{2.2in}
			\centering
			\includegraphics[width=2.2in,height=1.6in]{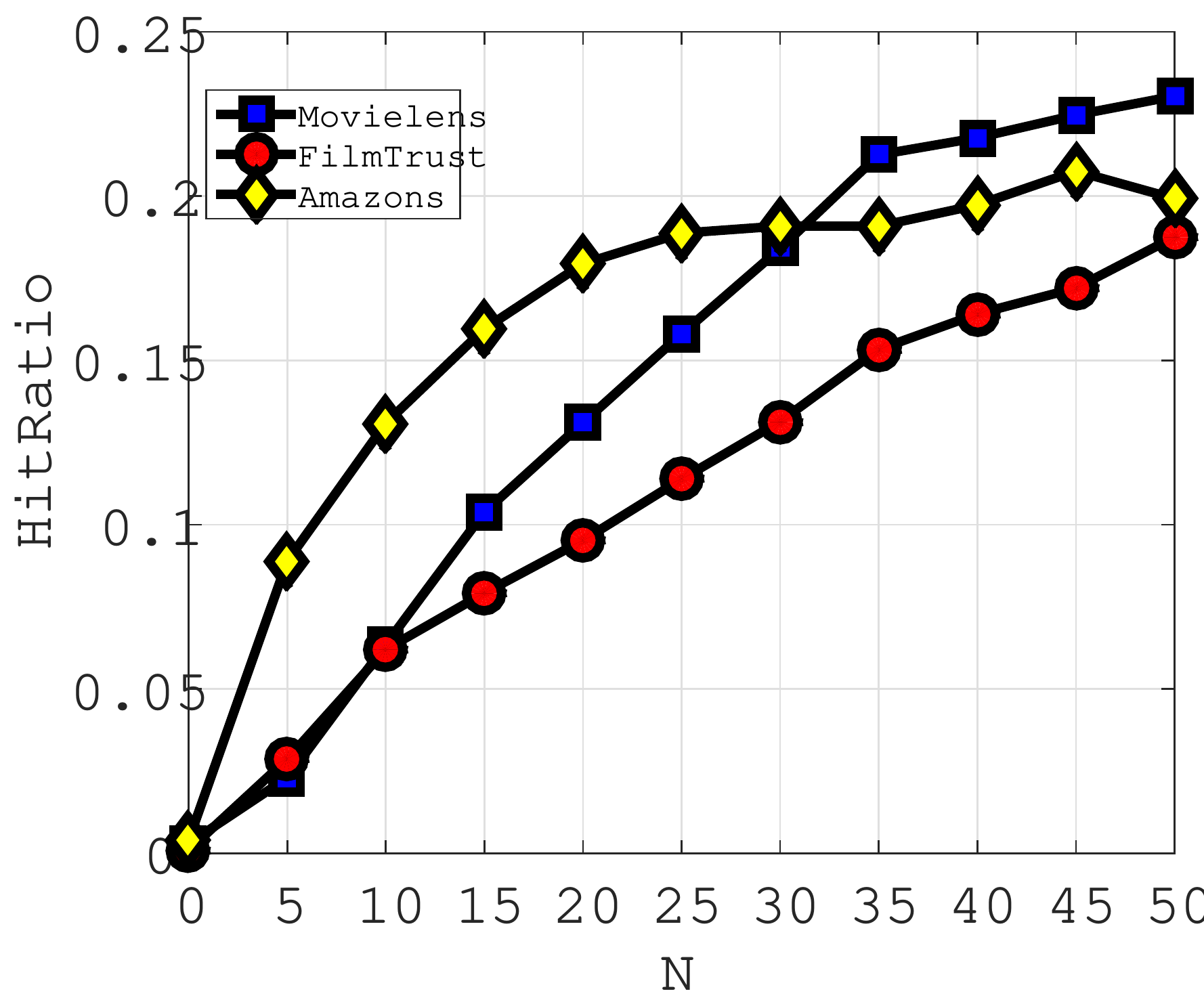}%
		\end{minipage}
	}
	\subfigure[]{
		\begin{minipage}{2.2in}
			\centering
			\includegraphics[width=2.2in,height=1.6in]{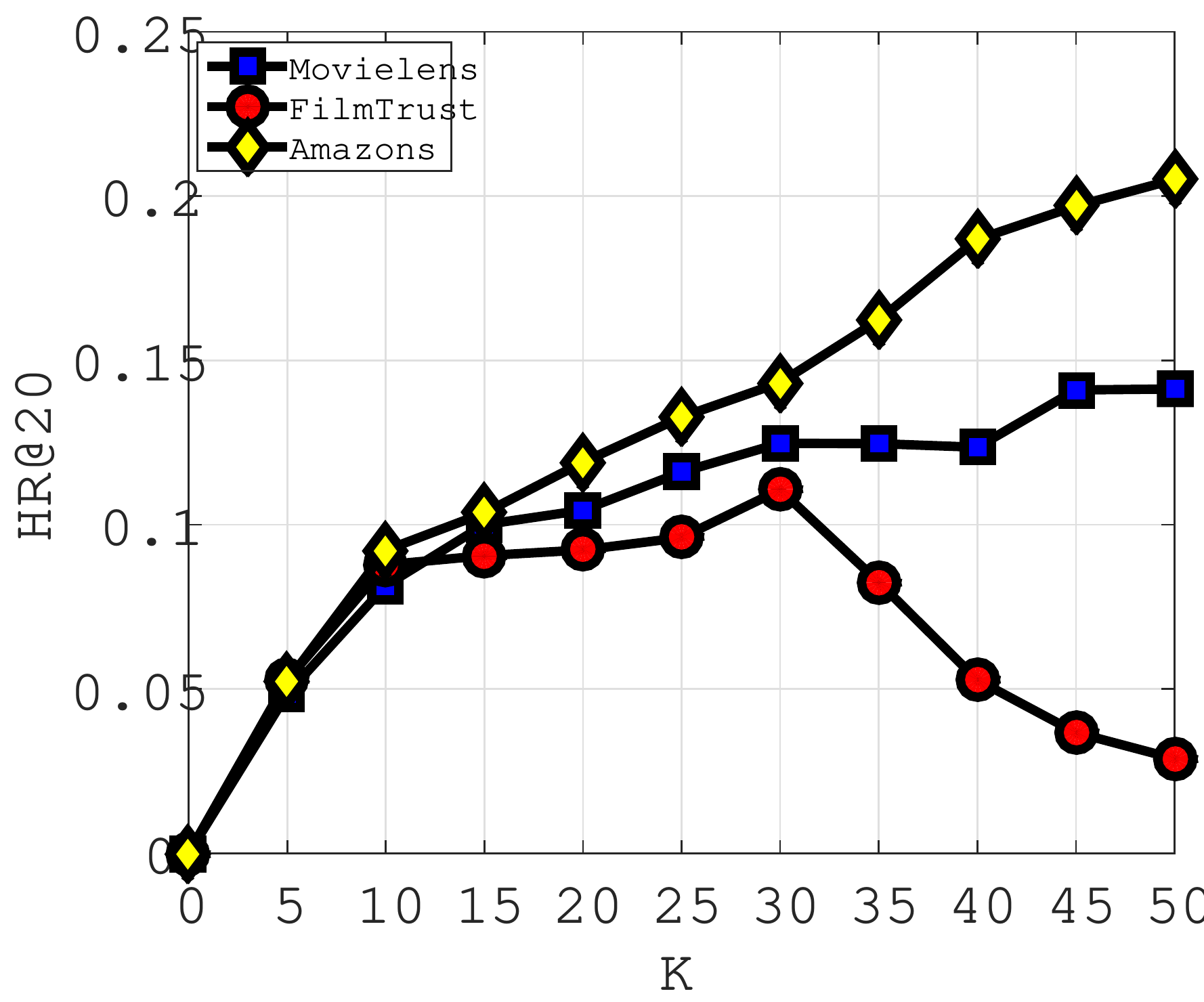}%
		\end{minipage}
	}
	\subfigure[]{
		\begin{minipage}{2.2in}
			\centering
			\includegraphics[width=2.2in,height=1.6in]{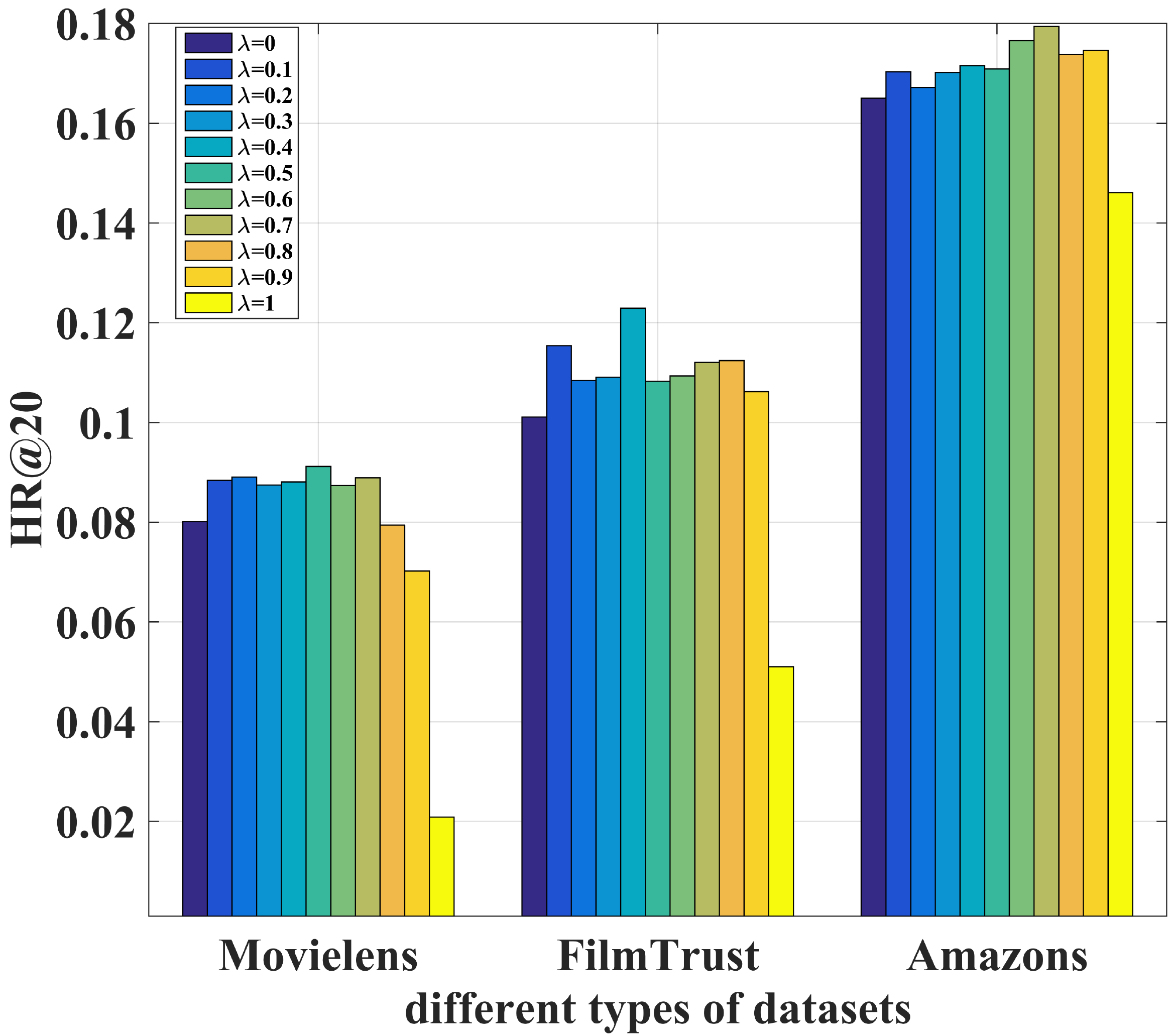}%
		\end{minipage}
	}
	
	\caption{Impact of parameters: the number of filler-items, the number of recommended items, the number of users' nearest neighbours and the weight of different loss functions.}
	\label{fig2}
\end{figure*}

\textbf{Impact of the number of filler-items.} Figure \ref{fig2} (a), (b) and (c) show the impact of the number of filler-items $z$ on our attack method, where the attack size = $2\%$ and we attack random target items. As we can see, in FilmtTrust and Amazons datasets when $z$ is close to the average rating number of normal users (23 in FilmTrust, 8 in Amazons), the attack can achieve better performance. If $z$ is larger than the average rating number, it is more difficult for fake users to imitate the behaviour of real users, which decreases the performance of similarity calculation between fake users and real users.  In the dense dataset Movielens, it can be clearly seen that our attack method achieves more competitive and stable performances if $z$ is set higher than 50.

\textbf{Impact of the number of recommended items.} We fix the attack size = $2\%$, attack random target items and set the number of recommended items $N $ to $\left\{1,5,10,15,20,25,30,35,40,45,50 \right\}$. In Figure \ref{fig2} (d), the smaller $N$ represents the harder the neighborhood-based recommender system is to be attacked. From the figure, we can find that when $N$ is small, our attack method can also achieve good performance. For instance, in Amazons dataset, when $N = 5$, our method can still obtain 0.089 $HitRatio$ (close to  the $HitRatio$ of warm-start items), which means our method can effectively promote the target items ranked in top-5 recommendation list.

\textbf{Impact of the number of users' nearest neighbours.} In Figure \ref{fig2} (e),  the number of  nearest neighbours of users $K$ is set to $\{1, 5, 10, 15, 20, 25, 30, 35, 40, 45, 50\}$, where the attack size = $2\%$ and we attack random target items. In FilmtTrust and Amazons datasets, we can observe that the $HitRatio$ increases as user-based CF considers more nearest neighbours of users. For instance, when the fake user is the 25th nearest neighbour of user $u$, if user $u$ only considers the top-20 nearest neighbours, he can avoid the attack; if the user considers top-30 nearest neighbours, he will be effected. However, in the Movielens, we notice that the $HitRatio$ decreases after $K$ is set higher than 30. The possible reason is that Movielens is a dense dataset. Setting higher $K$ for user $u$ may contain more normal users who have rated the target items, thus pushing the fake users out of top-K nearest neighbours.

\textbf{Impact of the weight of different loss functions.} Figure \ref{fig2} (f) shows the performance of our attack method depending on the value of $\lambda$, where the attack size = $2\%$ and we attack random target items. $\lambda$ controls the relative importance of $ loss_{1}$  and $loss_{2} $. Setting $\lambda$ to 1 or 0 means we only consider the $loss_{1}$ or $ loss_{2}$ respectively in the overall loss function. It can be noticed that the $loss_{1}$ imposes higher impact on the performance of our method. Besides, we can see that our attack method in different datasets achieves the best performance with different $\lambda$ values (e.g. $\lambda$ = 0.6 for FilmTrust, $\lambda$ = 0.5 for Movielens, $\lambda$ = 0.3 for Amazons).

\subsection{Transferability}

\setlength{\parindent}{2em}
Previous sections demonstrate the performance of our method in the white-box setting. In addition, adversarial samples usually have another important property, namely transferability \cite{papernot2016transferability}. Adversarial samples that are generated based on a certain model can also successfully fool other models with the same task. This case of attacks is named black-box attacks. Nowadays, web and apps show and collect information through a terminal machine. Then, a model deployed on cloud servers will handle this information. In this case, an attacker is able to acquire data through crawlers, but the structure and parameters of the model are hard to acquire because the model is deployed on cloud servers. Therefore, we consider that data poisoning attack in the black-box setting is generally more realistic and common than in the white-box setting.  

To explore the performance of our attack method in the black-box setting, we inject fake users generated by UNAttack to against other CF recommender systems and explore the robustness of them. Two state-of-the-art CF recommender systems, i.e., BPRMF\cite{rendle2009bpr} and NCF\cite{he2017neural}, are chosen as illustrative examples. Firstly, the fake user will be generated  based on the neighborhood-based recommender system. In this experiment, K and N are set to 30 and 20 respectively, and the number of filler-items is set to the average rating number of normal users. Note that in this experiment, the Cosine Similarity is chosen in training the fake user's phrase. In fact, other similarity measurement methods also can be applied such as Euclidean Distance-based Similarity and Pearson Similarity. Secondly, for the model BPRMF and NCF, we fix the embedding size to 20 and N to 20, and the learning rate is set to $\left\{0.001,0.005,0.01\right\}$. The original $HitRatio$ of target items will be calculated after the training based on the original data $R_{m \times n}$. We define the $R_{m \times n}$ added with fake users as $R_{m \times n}^{'}$ and train the model based on the $R_{m \times n}^{'}$ to obtain the new $HitRatio$.

\begin{table*}[!htbp]
	\centering
	\setlength{\tabcolsep}{2.0mm}
	\caption{Overall performance of transferability.}
	\begin{tabular}{|l|l|c|c|c|c|c|c|}
		\hline
		\multirow{2}{*}{Dataset}   & \multirow{2}{*}{Method} & \multicolumn{3}{c|}{BPRMF}                          & \multicolumn{3}{c|}{NCF}                            \\ \cline{3-8} 
		&                         & 0.50\%          & 1\%             & 2\%             & 0.50\%          & 1\%             & 2\%             \\ \hline
		\hline
		\multirow{6}{*}{Filmtrust} & None                    & 0.0012          & 0.0012          & 0.0012          & 0.0011          & 0.0011          & 0.0011          \\ \cline{2-8} 
		& Random                  & 0.0055          & 0.0071          & 0.0079          & 0.0087          & 0.0098          & 0.0312          \\ \cline{2-8} 
		& Average                 & 0.0055          & 0.0071          & 0.0079          & 0.0087          & 0.0098          & 0.0317          \\ \cline{2-8} 
		& Popular                 & 0.0055          & 0.0071          & 0.0079          & 0.0087          & 0.0097          & 0.0312          \\ \cline{2-8} 
		& Co-visitation             & 0.0162          & 0.1939          & 0.2164          & 0.0243          & 0.0854          & 0.0981          \\ \cline{2-8} 
		& UNAttack                & \textbf{0.0942} & \textbf{0.2266} & \textbf{0.2565} & \textbf{0.0668} & \textbf{0.2042} & \textbf{0.2223} \\ \hline
		\hline
		\multirow{6}{*}{Movielen}  & None                    & 0.0052  
       & 0.0052          & 0.0052          & 0.0079          & 0.0079          & 0.0079          \\ \cline{2-8} 
		& Random                  & 0.0052          & 0.0128          & 0.0173          & 0.0079          & 0.0085          & 0.0154          \\ \cline{2-8} 
		& Average                 & 0.0052          & 0.0129          & 0.0176          & 0.0079          & 0.0087          & 0.0154          \\ \cline{2-8} 
		& Popular                 & 0.0052          & 0.0129          & 0.0175          & 0.0079          & 0.0084          & 0.0158          \\ \cline{2-8} 
		& Co-visitation             & 0.0082          & 0.0130          & 0.0189          & 0.0079          & 0.0090          & 0.0169          \\ \cline{2-8} 
		& UNAttack                & \textbf{0.0115} & \textbf{0.0159} & \textbf{0.0247} & \textbf{0.0113} & \textbf{0.0184} & \textbf{0.0247} \\ \hline
		\hline
		\multirow{6}{*}{Amazons}   & None                    & 0.0033          & 0.0033          & 0.0033          & 0.0055          & 0.0055          & 0.0055          \\ \cline{2-8} 
		& Random                  & 0.0174          & 0.0874          & 0.0132          & 0.0142          & 0.0157          & 0.0652          \\ \cline{2-8} 
		& Average                 & 0.0176          & 0.0875          & 0.0131          & 0.0141          & 0.0153          & 0.0651          \\ \cline{2-8} 
		& Popular                 & 0.0172          & 0.0867          & 0.0132          & 0.0142          & 0.0153          & 0.0653 
		\\ \cline{2-8} 
		& Co-visitation             & 0.0562          & 0.2890          & 0.3443          & 0.0981          & 0.1519          & 0.1657          \\ \cline{2-8} 
		& UNAttack                & \textbf{0.0993} & \textbf{0.3651} & \textbf{0.4082} & \textbf{0.1167} & \textbf{0.2946} & \textbf{0.3260} \\ \hline
	\end{tabular}
	\label{tab3}
\end{table*}

From the experimental results shown in Table \ref{tab3}, we can obtain the following observations:

\begin{itemize}
	\item As can be observed, all data poisoning attack methods can be transferred to attack both BPRMF and NCF, and if attackers inject enough fake users into recommender systems, they can improve the $HitRatio$ upon the case of None. For example, in the FilmTrust dataset, when attacking BPRMF and NCF by injecting $2\%$ fake users, the HitRatio of UNAttack achieves 0.2565 and 0.2223, which is 213 times and 199 times higher than the None. This phenomenon can be explained by the fact that almost all the CF recommender systems will learn the correlation between users' historical information, which will be utilized to generate a personalized ranking list for users. The correlation learned by different CF recommender systems may be similar. Therefore, the fake users generated by UNAttack also can be transferred to attack BPRMF and NCF effectively. 
	\item  In the dense dataset, it is harder for the attacker to achieve the improvement of $HitRatio$. For example, when the attack size is set to $0.5\%$, in the Movielens dataset,  UNAttack only improves $HitRatio$ 4.7 times over the None lower than 15 times in the white-box setting. This phenomenon is similar in attacking the neighborhood-based CF.
	\item  Among all methods, UNAttack largely outperforms the traditional attack methods and the state-of-the-art methods Co-visitation whether attacking the BPRMF or NCF on the three datasets. For instance, in the Amazons dataset, when attacking NCF by injecting $2\%$ fake users, the best HitRatio of Co-visitation \cite{yang2017fake} reaches 0.1656 and our UNAttack can obtain 0.3260. This roughly 1 times relative improvement demonstrates the advantage of combining optimal technique with data poisoning. 
	\item NCF has better robustness against the data poisoning method than BPRMF. For example, in the Amazons dataset, when attacking NCF by injecting $2\%$ fake users, UNAttack achieves 59.4 times improvement compared with the None, whereas it can achieve 122.8 times improvement in attacking BPRMF. The reason lies in that NCF unifies the linearity of MF and the non-linearity of multi-layer perceptron (MLP). 
\end{itemize}

\section{Conclusion}

\setlength{\parindent}{2em}

This paper proposes a novel target attack framework against recommmender systems. Data poisoning attack against neighborhood-based methods is encoded into our framework from the perspective of optimization. The generated fake users are injected into original data when attacking the neighborhood-based recommendation so that we can improve the $HitRatio$ of the target items. 
Further, a cold-start item can be changed to a warm-start item. In addition, attack on BPRMF and NCF are conducted to demonstrate that our method can work in the black-box setting.
The results of the experiments on three real-world datasets indicate the effectiveness and transferability of our method via comparing with the state-of-the-art methods when targeting different types of target items.

In the future, how to achieve the good performance of our method will be studied, even if we only know part of the dataset. With the rapid development of deep learning based
recommender systems, we would like to design data poisoning attacks against the recommender systems based on deep learning. Moreover, we are interested in building an effective defence model to detect the data poisoning attack.

\section*{Acknowledgments}
This work supported by the National Key Research and Development Program (2017YFB0202201), the National Natural Science Foundation of China (61702568,U1711267), the Program for Guangdong Introducing Innovative and Entrepreneurial Teams (2017ZT07X355) and the Fundamental Research Funds for the Central Universities under Grant (17lgpy117).

\bibliography{wileyNJDAMA}%




\end{document}